\documentclass[a4paper]{article}

\usepackage{INTERSPEECH2020}

\usepackage{subcaption}
\usepackage{enumitem}

\title{MultiSpeech: Multi-Speaker Text to Speech with Transformer}
\name{Mingjian Chen$^1$\thanks{This work was done while the first, fourth and fifth authors were interning at Microsoft. Correspondence to: Tao Qin.}, Xu Tan$^2$, Yi Ren$^3$, Jin Xu$^4$, Hao Sun$^1$, Sheng Zhao$^5$, Tao Qin$^2$, Tie-Yan Liu$^2$}
\address{
  $^1$School of Software and Microelectronics, Peking University\\
  $^2$Microsoft Research Asia, $^3$Zhejiang Univeristy, $^4$Tsinghua University, $^5$Microsoft Azure Speech}
\email{milk@pku.edu.cn, xuta@microsoft.com, rayeren@zju.edu.cn, j-xu18@mails.tsinghua.edu.cn, sigmeta@pku.edu.cn, Sheng.Zhao@microsoft.com, taoqin@microsoft.com, tyliu@microsoft.com}

\begin{document}

\maketitle
\begin{abstract}
Transformer-based text to speech (TTS) model (e.g., Transformer TTS~\cite{li2019neural}, FastSpeech~\cite{ren2019fastspeech}) has shown the advantages of training and inference efficiency over RNN-based model (e.g., Tacotron~\cite{shen2018natural}) due to its parallel computation in training and/or inference. However, the parallel computation increases the difficulty while learning the alignment between text and speech in Transformer, which is further magnified in the multi-speaker scenario with noisy data and diverse speakers, and hinders the applicability of Transformer for multi-speaker TTS. In this paper, we develop a robust and high-quality multi-speaker Transformer TTS system called MultiSpeech, with several specially designed components/techniques to improve text-to-speech alignment: 1) a diagonal constraint on the weight matrix of encoder-decoder attention in both training and inference; 2) layer normalization on phoneme embedding in encoder to better preserve position information; 3) a bottleneck in decoder pre-net to prevent copy between consecutive speech frames. Experiments on VCTK and LibriTTS multi-speaker datasets demonstrate the effectiveness of MultiSpeech: 1) it synthesizes more robust and better quality multi-speaker voice than naive Transformer based TTS; 2) with a MutiSpeech model as the teacher, we obtain a strong multi-speaker FastSpeech model with almost zero quality degradation while enjoying extremely fast inference speed.

\end{abstract}

\section{Introduction}
In recent years, neural text to speech (TTS) models such as Tacotron~\cite{wang2017tacotron,shen2018natural}, Transformer TTS~\cite{li2019neural} and FastSpeech~\cite{ren2019fastspeech} have led to high-quality single-speaker TTS systems using large amount of clean training data. Thanks to the parallel computation in Transformer~\cite{vaswani2017attention}, Transformer based TTS enjoys much better training~\cite{li2019neural,ren2019fastspeech} and inference~\cite{ren2019fastspeech} efficiency than RNN based TTS~\cite{wang2017tacotron,shen2018natural}. 

To reduce deployment and serving cost in commercial applications, building a TTS system supporting multiple (hundreds or thousands) speakers has attracted much attention in both industry and academia~\cite{gibiansky2017deep,ping2017deep,shen2018natural,hsu2018hierarchical}. While it is affordable to record high-quality and clean voice in professional studios for a single speaker, it is costly to do so for hundreds or thousands of speakers to build a multi-speaker TTS system. Thus, multi-speaker TTS systems are usually built using multi-speaker data recorded for automatic speech recognition (ASR)~\cite{panayotov2015librispeech,zen2019libritts} or voice conversion~\cite{veaux2016superseded}, which is noisy and of low-quality due to the diversity and variances of prosodies, speaker accents, speeds and recording environments. Although Transformer based models have shown advantages over other neural models for single-speaker TTS, existing works on multi-speaker TTS mostly adopt RNN (e.g., Tacotron~\cite{wang2017tacotron,shen2018natural}) or CNN (e.g., Deep Voice~\cite{gibiansky2017deep,ping2017deep}) as the model backbone, and few attempts have been made to build Transformer based multi-speaker TTS.

The main challenge of Transformer multi-speaker TTS comes from the difficulty of learning the text-to-speech alignment, while such alignment plays an important role in TTS modeling~\cite{shen2018natural,ping2017deep,ren2019fastspeech}. While applying Transformer to multi-speaker TTS, the text-to-speech alignment between the encoder and decoder is more difficult than that of RNN models. When calculating the attention weights in each decoder time step in RNN, advanced strategies such as location-sensitive attention~\cite{chorowski2015attention} are leveraged to ensure the attention move forward consistently through the input, avoiding word skipping and repeating problems. Location-sensitive attention leverages the attention results in previous decoder time steps, which, unfortunately, cannot be used in Transformer due to parallel computation during training. In single-speaker TTS, the text and speech data are usually of high-quality and the text-to-speech alignments are easy to learn. However, as aforementioned, the speech data for multi-speaker TTS is usually noisy, which makes the alignments much more difficult. Actually, CNN multi-speaker TTS also faces this challenge, and complex systems are designed based on the characteristics of CNN structure in~\cite{gibiansky2017deep,ping2017deep}, which unfortunately cannot be easily applied on Transformer models.

\begin{figure}[!t]
    \centering
    \setlength{\belowcaptionskip}{-0.5cm} 
    \includegraphics[width=0.5\textwidth]{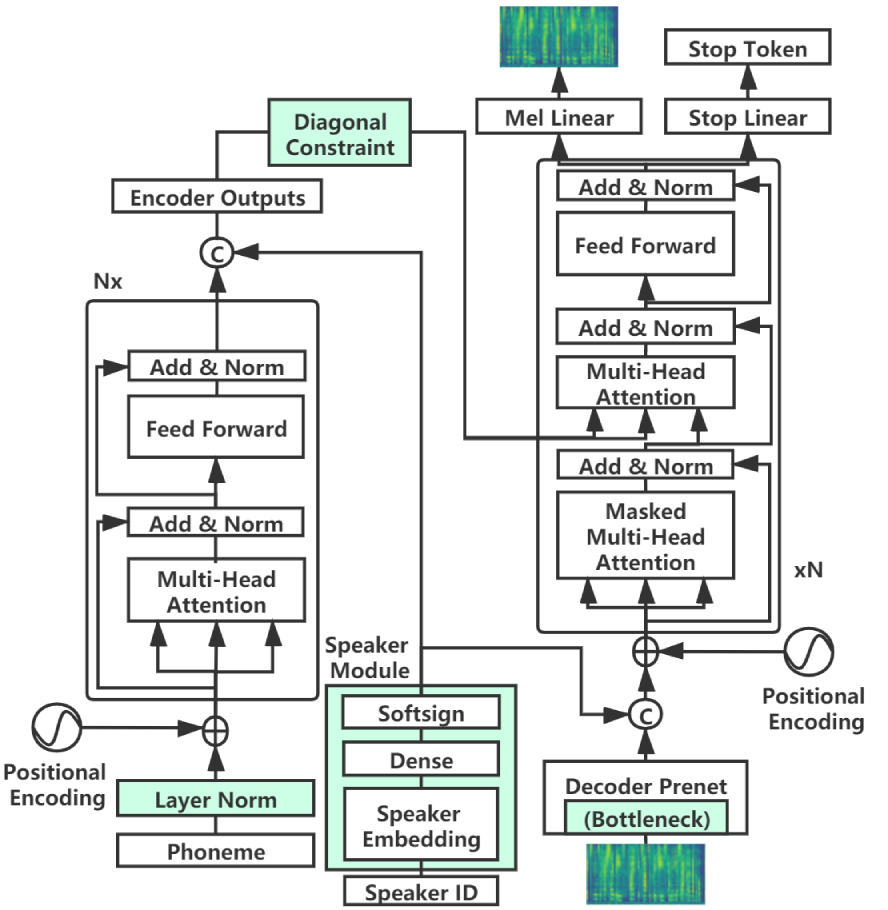}  
    \caption{The model structure of our proposed MultiSpeech. The green blocks are the newly added modules for multi-speaker TTS based on Transformer.}
    \label{fig_multispeech}
\end{figure}

In order to bring the advantages of Transformer into multi-speaker TTS modeling, in this paper, we develop a robust and high-quality multi-speaker TTS system called MultiSpeech, which greatly improves the text-to-speech alignment in Transformer. Specifically, we introduce several techniques to improve the alignments based on empirical observations and insights. First, considering the attention alignments between the text encoder and speech decoder are usually monotonic and diagonal, we introduce a diagonal constraint on the weight matrix of the encoder-decoder attention during training and inference. Second, position embeddings are important in Transformer and can help text-to-speech alignment~\cite{ping2017deep}\footnote{Ideally, the model can learn the monotonic alignment simply through the position embeddings in text and speech sequences.}, and are usually added to phoneme embeddings in the Transformer encoder. However, the scale of phoneme embeddings can vary a lot while that of position embeddings is fixed, which causes magnitude mismatch\footnote{The embeddings of some phonemes are large and will dominate position embeddings, and some phonemes are of small embeddings and will be dominated by position embeddings, both of which will harm the alignment learning.} while added together and consequently increases the difficulty of model training. Therefore, we add a layer normalization step on phoneme embeddings to make them comparable and better preserve position information. Third, text-to-speech alignments should be learnt by attending to source phonemes while generating target speech frames. However, two adjacent speech frames are usually similar and standard Transformer decoder tends to directly copy previous frame to generate the current frame. Consequently, no alignments between text and speech can be learned. To prevent direct copy between consecutive speech frames, we employ a bottleneck structure in the decoder pre-net which encourages the decoder to generalize on the representation of speech frame instead of memorization, and forces the decoder to attend to text/phoneme inputs.

Experiments on VCTK and LibriTTS multi-speaker datasets show that 1) MultiSpeech achieves great improvements (1.01 MOS gain on VCTK and 1.46 MOS gain on LibriTTS) over naive Transformer based TTS and synthesizes robust and high-quality multi-speaker voice. 2) The three proposed techniques can indeed improve text-to-speech alignments, measured by the attention diagonal rate. 3) A well trained MultiSpeech model can be used as a teacher for FastSpeech training and we obtain a strong multi-speaker FastSpeech model without quality degradation but enjoying extremely fast inference.

\section{Background}
\textit{Transformer TTS.}
Transformer based TTS (e.g.~\cite{li2019neural}) adopts the basic model structure of Transformer~\cite{vaswani2017attention}, as shown in Figure~\ref{fig_multispeech} (remove the green blocks). Each transformer block consists of a multi-head self-attention network and a feed-forward network. Additionally, a decoder pre-net is leveraged to pre-process the mel-spectrogram frame, and a mel linear layer is used to predict the mel-spectrogram frame and a stop linear layer to predict if should stop in each predicted frame. Transformer can ensure parallel computation during training, which, as a side effect, harms the attention alignments between text and speech, as analyzed in the introduction part. As a result, it is challenging to build multi-speaker TTS on Transformer considering the complicated acoustic conditions in multi-speaker speech. In this paper, we analyze each component in Transformer TTS to figure out why it fails to learn alignments, and propose the corresponding modifications to improve the alignments.

\textit{Multi-Speaker TTS.} Several works have built multi-speaker text to speech systems based on RNN~\cite{jia2018transfer,skerry2018towards} and CNN~\cite{gibiansky2017deep,ping2017deep}. RNN-based multi-speaker model enjoys the benefits of recurrent attention computation as in Tacotron 2~\cite{shen2018natural}, which can leverage the attention information in previous steps to help the attention calculation in current step. CNN-based multi-speaker model~\cite{ping2017deep} develops many sophisticated mechanisms in the speaker embedding and attention block to ensure the synthesized quality. VAE-based method~\cite{hsu2018hierarchical} is further leveraged to handle noisy multi-speaker speech data~\cite{zen2019libritts}. Considering the advantages of Transformer including parallel training over RNN and effective sequence modeling over CNN, in this paper, we build multi-speaker TTS on Transformer model.

\textit{Text-to-Speech Alignment.} Since text and speech correspond to each other in TTS, the alignments between text and speech are generally monotonic and diagonal in the encoder-decoder attention weights. Previous works have tried different techniques to ensure the alignments between text and speech in encoder-decoder model. Location sensitive attention~\cite{chorowski2015attention} is proposed to align the source and target better by leveraging previous attention information. ~\cite{he2019robust,zhang2018forward,tachibana2018efficiently,yasuda2019initial} improve text-to-speech alignments by designing sophisticated techniques on attention. ~\cite{ping2017deep} design position encoding with its angular frequency determined dynamically by each speaker embedding to ensure the text-speech alignment. \cite{wang2017tacotron} uses large dropout in decoder pre-net and finds it is helpful for attention alignment. In this paper, we introduce several techniques to improve the alignments specifically in Transformer model.

\section{Improving Text-to-Speech Alignment}
In this section, we introduce several techniques to improve the text-to-speech alignments in MultiSpeech, from the attention, encoder and decoder part respectively, as shown in Figure~\ref{fig_multispeech}.



\begin{figure}[!t]
    \centering
       \setlength{\belowcaptionskip}{-0.5cm}  \includegraphics[width=0.4\textwidth]{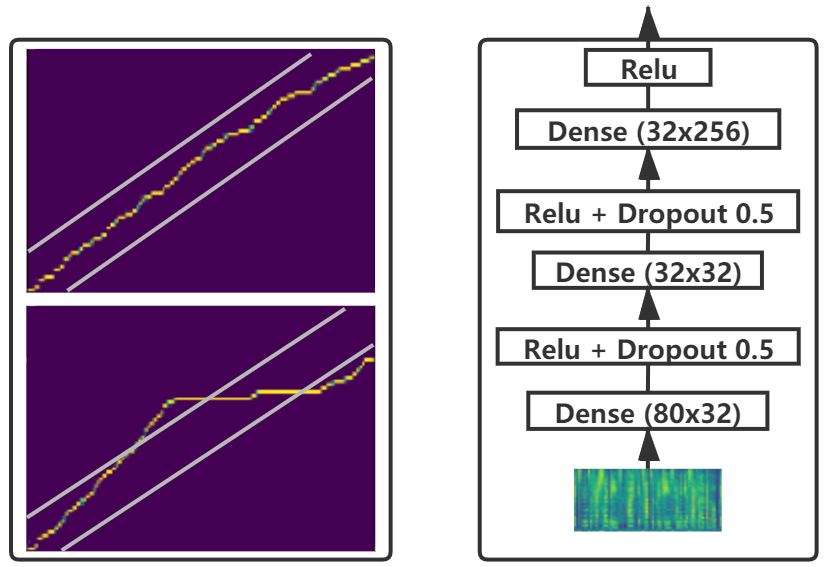}
\begin{tabular}{ll}
  \hspace{0.3cm}(a)  &
  \hspace{2.7cm}(b) 
\end{tabular}
\caption{ (a) The illustration of diagonal constraint in attention, where the above figure has a small diagonal constraint loss and the below figure has a large diagonal constraint loss. (b) The model structure of the pre-net bottleneck in decoder.}
\label{fig_constraint_bottleneck}
\end{figure}

\subsection{Diagonal Constraint in Attention}
\label{sec_dc}
Monotonic and diagonal alignments in the attention weights between text and speech are critical to ensure the quality of synthesized speech~\cite{chorowski2015attention,he2019robust,zhang2018forward,ping2017deep,tachibana2018efficiently,yasuda2019initial}. In multi-speaker scenario, the speech is usually noisy and different speakers have different speeds and acoustic conditions, making the alignments difficult. Therefore, we propose to add diagonal constraint on the attention weights to force the model to learn correct alignments. 

We first formulate the diagonal attention rate $r$ as 
\begin{equation}
\setlength{\abovedisplayskip}{3pt}
\setlength{\belowdisplayskip}{3pt}
r = \frac{\sum_{t=1}^{T} \sum_{s=kt-b}^{kt+b} A_{t,s}} {S},
\label{eqn_r}
\end{equation}
where $S$ is the length of speech mel-spectrogram and $T$ is the length of text (phoneme or character). $k=\frac{S}{T}$ is the slop for each training sample and $b$ is a hyperparameter for bandwidth, both of which determine the shape the diagonal area. $A_{t,s}$ is the $t$-th row and $s$-th column of the attention weight matrix $A$. The numerator represents how much weight lie in the diagonal area while the denominator represents the total attention weight which equals to speech length $S$. The diagonal constraint loss $L_{DC}$ encourages larger attention weights in the diagonal area as shown in Figure~\ref{fig_constraint_bottleneck}, which is defined as $L_{DC} = - r$, 
where $r$ is defined in Equation~\ref{eqn_r}. $L_{DC}$ is added on the original TTS loss with a weight $\lambda$ to adjust the strength of the constraint.

In order to ensure the correct alignment during inference, we also add attention constrain in the autoregressvie generation process. We introduce an attention sliding window in the text side and compute the attention weights only within this window. The range of the window is [-1, 4], where 0 in the window represents the window center and is initialized as position 0 in the beginning. The window allows the predicted frame to attend on both previous 1 phoneme and future 4 phonemes of the center. We design a sliding window moving strategy: we define the attention centroid of $s$-th predicted frame as $C_s = \lfloor \sum_{t=0}^{T}(A_{t,s}*t) \rfloor $. If $C_s$ deviates the window center beyond 3 consecutive frames, we move the sliding window center one step forward.

Compared with the attention constraint strategy proposed in~\cite{ping2017deep}, our method has the following advantages: 1) our sliding window allows to attend to the previous position, and 2) we use attention centroid rather than simply the position of the highest attention weight within the current window as new sliding window center. These improvements can prevent the sliding window from moving forward too early, which usually results in skipping phonemes and fast speaking speed.

\subsection{Position Information in Encoder}
The encoder of Transformer based TTS model usually takes $x+p$ as input, where $x$ is the embedding of phoneme/character token and $p$ is positional embedding to give the Transformer model a sense of token order. $p$ is usually formulated as triangle positional embeddings~\cite{vaswani2017attention} and the scale of its value is fixed into $[-1, 1]$. However, the embedding $x$ is learned end-to-end, and the scale of the its value can be very large or small. As a result, the position information $p$ in $x+p$ is relatively small or large, which will affect the alignment learning between the source (text) and target (speech) sequence. 

To preserve the position information properly in $x+p$, we first add layer normalization~\cite{ba2016layer} on $x$ and then add with $p$, i.e., $LN(x)+p$, as shown in Figure~\ref{fig_multispeech}. $LN(x)$ is defined as 
\begin{equation}
\setlength{\abovedisplayskip}{3pt}
\setlength{\belowdisplayskip}{3pt}
LN(x) = \gamma \frac{x-\mu}{\sigma} + \beta,
\end{equation}
where $\mu$ and $\sigma$ are the mean and variance of vector $x$, $\gamma$ and $\beta$ are the scale and bias parameters. In this case, the scale of phoneme embedding $x$ can be restricted to a limited range by learning the scale and bias parameters in layer normalization. 

In Transformer TTS~\cite{li2019neural}, a scalar trainable weight $\alpha$ is leveraged to adjust $p$ before adding on $x$, i.e., $x+\alpha p$. However, it cannot necessarily ensure enough position information in $x+\alpha p$, since a single scalar $\alpha$ cannot balance the scales between position information $p$ and embedding $x$, considering different phonemes/characters have different scales\footnote{We do not normalize the input in decoder, since mel-specotrgram is not learnable and usually normalized into a fixed range. This point is also confirmed in~\cite{li2019neural}, where the scalar trainable weight in decoder is much more stable and closer to 1 than that in encoder.}. We also verify the advantage of our layer normalization over simple scalar trainable weight in the experiment part\footnote{Our layer normalization is also better than learnable position embeddings since it still learns a global embedding for each position.}.

\subsection{Pre-Net Bottleneck in Decoder}
The adjacent frames of mel-spectrogram are usually very similar since the hop size is usually much smaller than the window size\footnote{The typical parameters of window size and hop size in TTS is 50ms and 12.5ms.}, which means two adjacent frames have large information overlap. As a consequence, when predicting next frame given current frame as input in autoregressive training, the model is prone to directly copy some information from the input frame instead of extracting information from text side for meaningful prediction. The decoder pre-net in~\cite{wang2017tacotron} leverages a structure like 80-256-128 where each number represents the hidden size of each layer in the pre-net, while the decoder pre-net in~\cite{shen2018natural,li2019neural} leverages a structure like 80-256-256-512, both with dropout rate of 0.5. The authors~\cite{wang2017tacotron,shen2018natural} claim this structure can act like a bottleneck to prevent from copy (the hidden size is halved in the bottleneck, e.g., 128 vs.256 or  256 vs. 512). However, the mel-spectrogram with a dimension of 80 is first converted into 512 or 256 hidden and is then halved to 256 or 128, which is still larger than 80 and cannot necessarily prevent copy and learn alignments in multi-speaker scenario, according to our experiments. As shown in Figure~\ref{fig_constraint_bottleneck}, we further reduce the bottleneck hidden size to as small as $1/8$ of the original hidden size (e.g., 32 vs. the original hidden size 256) plus with 0.5 dropout ratio, and the structure becomes 80-32-32-256. We found this small bottleneck size is essential to learn meaningful alignments and avoid direct copying input frame.

\section{Experiments and Results}
In this section, we conduct experiments to verify the advantages of MultiSpeech and the effectiveness of the proposed techniques to improve text-to-speech alignments. 

\subsection{Experimental Setup}
\textit{Datasets.} We conducted experiments on the VCTK~\cite{veaux2016superseded} and LibriTTS~\cite{zen2019libritts} multi-speaker datasets. The VCTK dataset contains 44 hours speech with 108 speakers, while the LibriTTS dataset contains 586 hours speech with 2456 speakers. We convert the speech sampling rate of both corpus to 16KHz, and use 12.5ms hop size, 50ms window size to extract mel-spectrogram. We convert text into phoneme using grapheme-to-phoneme conversion~\cite{sun2019token} and take phoneme as the encoder input.

\textit{Model Configuration.} The model structure of MultiSpeech is shown in Figure~\ref{fig_multispeech}. Both the encoder and decoder use 4-layer transformer blocks. The hidden size, attention head, feed-forward filter size and kernel size are 256, 2, 1024 and 9 respectively. In addition, the decoder pre-net bottleneck, as shown in Figure~\ref{fig_constraint_bottleneck}, is 32, which is 1/8 of the hidden size. For the speaker module as shown in Figure~\ref{fig_multispeech}, we follow the structure in~\cite{ping2017deep}.

\textit{Training and Inference.} We use 4 P100 GPUs, each with batch size of about 20,000 speech frames. We use Adam optimizer with $\beta_1 = 0.9$, $\beta_2 = 0.98$, $\epsilon = 10^{-9}$ and follow the learning rate schedule in~\cite{vaswani2017attention}. The bandwidth $b$ in the attention constraint is set to 50, and the weight $\lambda$ of $L_{DC}$ is set to 0.01 according to the valid performance. During inference, we use attention constraint as described in Section~\ref{sec_dc} to ensure the text-to-speech alignments. WaveNet~\cite{oord2016wavenet} is used as vocoder to synthesize voice. 

\textit{Evaluation.} We use MOS (mean opinion score) to measure the voice quality. Each sentence is judged by 20 native speakers. We also use the diagonal attention rate $r$ as defined in Equation~\ref{eqn_r} to measure the quality of text-to-speech alignments. A higher MOS means better voice quality while a higher $r$ means better alignments, and they are correlated to each other. For both VCTK and LibriTTS, we select 6 speakers (3 men and 3 women, each with 5 sentences) for evaluation respectively. 

\subsection{The Quality of MultiSpeech}
The MOS results are shown in Table~\ref{tab_mos_vctk}. We compare our proposed \textit{MultiSpeech} with 1) \textit{GT}, the ground-truth recording, 2) \textit{GT mel + Vocoder}, we first convert the recording into mel-spectrogram and then convert the mel-spectrogram back to audio with Vocoder, and 3) \textit{Transformer based TTS}, we only add the speaker embedding module on naive Transformer based TTS model to support multiple speakers, without using any of our proposed techniques to improve alignments. It can be seen that MultiSpeech achieves large MOS score improvements over Transformer based TTS. Transformer based TTS cannot learn effective alignments on most sentences and causes word skipping and repeating issues, or totally crashed voice. The MOS score of MultiSpeech on VCTK is also close to \textit{GT mel + Vocoder}. These results demonstrate the advantages of MultiSpeech for multi-speaker TTS. We show some demo audios and case analyses in this link\footnote{https://speechresearch.github.io/multispeech/}.

\begin{table}[th]
\setlength{\abovecaptionskip}{0.15cm} 
\setlength{\belowcaptionskip}{-5pt}
  \caption{The MOS scores with $95\%$ confidence intervals on VCTK and LibriTTS.}
  \label{tab_mos_vctk}
  \centering
  \begin{tabular}{ l c c }
    \toprule
    \textbf{Setting} & \textbf{VCTK} & \textbf{LibriTTS} \\
    \midrule
    \textit{GT} & $4.04\pm0.14$ & $4.14\pm0.16$\\
    \textit{GT mel + Vocoder} & $3.89\pm0.20$ & $3.90\pm0.08$\\
    \textit{Transformer based TTS} & $2.64\pm0.35$ & $1.49\pm0.09$\\
    \midrule
    \textit{MultiSpeech}  & $3.65\pm0.14$ & $2.95\pm0.14$\\
    \bottomrule
  \end{tabular}
  \vspace{-0.5cm}
\end{table}


\subsection{Method Analysis}
\textit{Ablation Study.} We first conduct ablation study on VCTK dataset to verify the effectiveness of each proposed technique: diagonal constraint (DC) in attention, layer normalization (LN) in encoder, pre-net bottleneck (PB) in decoder. The results are shown in Table~\ref{tab_ablation_study}. After removing diagonal constraint (DC), layer normalization (LN) and pre-net bottleneck (PB) respectively, both MOS score and diagonal rate $r$ drop. After further removing all the three techniques (-DC-LN-PB, i.e., Transformer based TTS), both MOS and $r$ drop largely. These ablation studies verify the effectiveness of the three techniques to improve attention alignments for better voice quality.

\begin{table}[th]
\setlength{\abovecaptionskip}{0.15cm}
\setlength{\belowcaptionskip}{-5pt}
  \caption{The MOS with $95\%$ confidence intervals and diagonal attention rate $r$ of the ablation study on VCTK. \textit{$-$DC} means not using diagonal constraint during training and inference. \textit{$-$LN} means using $x+p$ as encoder input but not $LN(x)+p$. \textit{$-$PB} means using pre-net structure like 80-256-256-256 instead of our proposed 80-32-32-256.}
  \label{tab_ablation_study}
  \centering
  \begin{tabular}{ l c c}
    \toprule
    \textbf{Setting} & \textbf{MOS} & \textbf{r} \\
    \midrule
     \textit{MultiSpeech} & $3.65\pm0.14$ & 0.694   \\    
  	 ~~\textit{$-$DC}        & $3.59\pm0.25$ & 0.502   \\    
     ~~\textit{$-$LN}        & $3.08\pm0.05$ & 0.637   \\
     ~~\textit{$-$PB}        & $3.36\pm0.27$ & 0.658   \\
     ~~\textit{$-$DC$-$LN$-$PB}  & $2.64\pm0.35$ &0.366   \\
    \bottomrule
  \end{tabular}
\vspace{-0.5cm}
\end{table}

\textit{Comparison between layer normalization and learnable weight.} We calculate the similarity between $p$ and three settings: 1) $LN(x)+p$, our proposed layer normalization (\textit{LN}); 2) $x+\alpha p$, the learnable weight (\textit{LW}) used in~\cite{li2019neural}; 3) $x+p$, the naive Transformer baseline (\textit{Baseline}). As shown in Table~\ref{tab_att_diagonal}, the similarity of \textit{LN} is in between \textit{LW} and \textit{Baseline}, which shows the position information in \textit{LN} is neither too weak (as in \textit{Baseline}) nor too strong (as in \textit{LW}\footnote{We check the final learnable weight $\alpha=2.62$, which is much bigger than the single-speaker setting in~\cite{li2019neural} ($\alpha$ is about $0.5$). We guess that when added with different scales of phoneme embedding, \textit{LW} simply learns a global large $\alpha$ to highlight position embedding.}) and is helpful for attention alignment. This is also verified by the diagonal attention rate $r$ in Table~\ref{tab_att_diagonal}. Our proposed \textit{LN} achieves the highest $r$ while \textit{LW} the lowest, which demonstrates that too strong position dominates phoneme embedding and harms the attention alignment.

\begin{table}[th]
\setlength{\abovecaptionskip}{0.15cm}
\setlength{\belowcaptionskip}{-5pt}
  \caption{The comparison of similarity and diagonal attention rate $r$ between \textit{LN}, \textit{LW} and \textit{Baseline} settings.}
  \label{tab_att_diagonal}
  \centering
  \begin{tabular}{ l c c c}
    \toprule
    \textbf{Setting} & \textit{LN} & \textit{LW} & \textit{Baseline}  \\
    \midrule
    \textbf{Similarity} & 0.126 & 0.184 & 0.089 \\ 
    \textbf{r} & 0.694 & 0.506 & 0.637  \\
    \bottomrule
  \end{tabular}
  \vspace{-0.5cm}
\end{table}

\subsection{Extension on FastSpeech}
We further use MultiSpeech as a teacher to teach a multi-speaker FastSpeech~\cite{ren2019fastspeech} on VCTK dataset, following the setting in~\cite{ren2019fastspeech}. We select 6 speakers (3 men and 3 women, each with 10 sentences) for MOS evaluation. As shown in Table~\ref{tab_mos_fastspeech}, we can obtain a strong FastSpeech model with nearly the same MOS score with MultiSpeech teacher\footnote{We can use more unlabeled text in knowledge distillation to further improve FastSpeech quality to match or even outperform teacher model.}.
\begin{table}[th]
\setlength{\abovecaptionskip}{0.15cm}
\setlength{\belowcaptionskip}{-5pt}
  \caption{The MOS score of multi-speaker FastSpeech on VCTK with $95\%$ confidence intervals.}
  \label{tab_mos_fastspeech}
  \centering
  \begin{tabular}{ l c c c}
    \toprule
    \textbf{Setting} & \textit{GT} & \textit{MultiSpeech} & \textit{FastSpeech}  \\
    \midrule
    \textbf{MOS}  & $4.02\pm0.09$ & $3.53\pm0.22$ & $3.45\pm0.13$ \\
    \bottomrule
  \end{tabular}
\end{table}
\vspace{-0.5cm}

\section{Conclusions}
In this paper, we developed MultiSpeech, a multi-speaker Transformer TTS system that leverages three techniques including diagonal constraint in attention, layer normalization in encoder and pre-net bottleneck in decoder, to improve the text-to-speech alignments in multi-speaker scenario. Experiments on VCTK and LibriTTS multi-speaker datasets demonstrate effectiveness of MutiSpeech: 1) it generates much higher-quality and more stable voice compared with Transformer TTS baseline; 2) using MultiSpeech as a teacher, we obtain a strong multi-speaker FastSpeech model to enjoy extremely fast inference speed. In the future, we will continue to improve the voice quality of MultiSpeech and multi-speaker FastSpeech model to deliver better multi-speaker TTS solutions.

\bibliographystyle{IEEEtran}

\bibliography{mybib}

\end{document}